\documentclass[conference, letterpaper]{IEEEtran}

\usepackage{mathtools, amssymb}
    \DeclareMathOperator*\Prob{Prob}
    \newtheorem{theorem}{Theorem}
    \newtheorem{lemma}[theorem]{Lemma}
    \newtheorem{proposition}[theorem]{Proposition}
    \def\loll{\begin{bmatrix} 1&0\\1&1 \end{bmatrix}}
    \def\cas#1{\begin{cases*} #1 \end{cases*}}

\usepackage{tikz-cd, pgfplots}
    \pgfplotsset{compat=1.18}
    \usetikzlibrary{calc}

\usepackage[colorlinks, allcolors=blue!50!purple!90!black]{hyperref}

\usepackage{autonum}

\begin{document}
                                 \title
               {Successive Cancellation Sampling Decoder:
         \\ An Attempt to Analyze List Decoding Theoretically}
                                    
                               \author
       {\IEEEauthorblockN{Hsin-Po Wang and Venkatesan Guruswami}
                           \IEEEauthorblockA
      {Department of Electrical Engineering and Computer Science
            \\ Simons Institute for the Theory of Computing 
        \\ University of California, Berkeley, Berkeley, CA, USA
             \\ Emails: \{simple, venkatg\}@berkeley.edu}}
                                    
                               \maketitle

\begin{abstract}\boldmath
    Successive cancellation list (SCL) decoders of polar codes excel in
    practical performance but pose challenges for theoretical analysis.
    Existing works either limit their scope to erasure channels or
    address general channels without taking advantage of soft
    information.  In this paper, we propose the \emph{successive
    cancellation sampling} (SCS) decoder.  SCS hires iid ``agents" to
    sample codewords using posterior probabilities.  This makes it fully
    parallel and amenable for some theoretical analysis.  As an example,
    when comparing SCS with $a$ agents to any list decoder with list
    size $\ell$, we can prove that the error probability of the former
    is at most $\ell/ae$ more than that of the latter.  In this paper,
    we also describe how to adjust the ``temperature'' of agents.
    Warmer agents are less likely to sample the same codewords and hence
    can further reduce error probability.
\end{abstract}

\section{Introduction}

    The \emph{Successive cancellation} (SC) decoder was invented
    alongside polar codes by Arıkan \cite{Ari09}.  Polar codes paired
    with SC enjoy beautiful theoretical properties such as achieving
    capacities of various channels.\footnote{ Notable examples include
    discrete memoryless channels \cite{STA09},
    multiple access channels \cite{MEL16},
    broadcast channels \cite{GAG15},
    wiretap channels \cite{MaV11},
    asymmetric channels \cite{HoY13},
    channels with memory \cite{GNS19},
    non-stationary channels \cite{Mah20}, and
    deletion channels \cite{TPF22}.}
    Later, Tal and Vardy \cite{TaV15} found that an \emph{SC list} (SCL)
    decoder is more powerful in practice, which is what the 5G standard
    \cite{BCL21} is currently using.  That said, SCL decoders are very
    difficult to analyze theoretically.

    Early analysis of SCL decoders are based on the observation
    that an erasure triggers a fork and doubles the list size $\ell$
    \cite[Theorem~4]{MHU15} \cite[Corollary~4]{CoP22}.  This observation
    falls apart over non-erasure channels due to soft bits.  Soft bits
    create dilemmas between
    (a) forking, just to protect a partially known bit, and
    (b) not forking, risking the selection of wrong paths.
    Some works choose option (a) \cite[Theorem~1]{HMH18}
    \cite[Theorem~1]{FVY21} and obtain achievable bounds that are very
    loose.  Others prove impossibility bounds \cite[Theorem~1]{MHU15}
    \cite[Theorem~1]{CoP22} that achievability bounds are unable to
    match unless we know exactly when to choose (b).

    The present paper attempt to bypass the difficulties by employing 
    ``soft-forking'' for general channels.  For general channels, the
    branches will be associated with posterior probabilities $p$ and
    $q$, among which the SC decoder will choose the greater one.  But
    here, we let an agent choose the $p$-branch with probability $p$ and
    the $q$-branch with probability $q$.  In other words, the agent will
    give the less favorable branch a chance from time to time.  We then
    hire $a$ independent agents to sample messages.  We call this an
    \emph{SC sampling} (SCS) decoder.

    Analyzing SCS is more straightforward than other list decoders:
    Instead of maintaining a list of prefixes, growing them into
    full-length vectors, and trying to characterize the highly dynamic
    nature of memory-swapping, we just let $a$ agents work in parallel
    and analyze them independently.  We hope that SCS bridges the gap
    between practical implementations and theoretical characterizations,
    unlocking new insights into the fundamental properties of polar
    codes and decoders.

    We present Theorem~\ref{thm:optimal} as a demonstration of what we
    want to achieve.

\begin{figure}
    \centering
    \begin{tikzpicture}[y=0.65cm]
        \draw
            (0, 0) coordinate (A) circle (1pt)
            node [above, align=center] {start \\ decoding}
            (-2, -1) coordinate (B1)
            (A) -- node [above, sloped] {49\%} (B1)
            node [below, align=center] {rejected by frozen bit}
            (2, -1) coordinate (B2) circle (1pt)
            (A) -- node [above, sloped] {51\%} (B2)
            (0, -2) coordinate (C1) circle (1pt)
            (B2) -- node [above, sloped] {38\%} (C1)
            (4, -2) coordinate (C2)
            (B2) -- node [above, sloped] {62\%} (C2)
            node [below, align=center] {rejected by \\ frozen bit}
            (-2, -3) coordinate (D1) circle (1pt)
            (C1) -- node[above, sloped] {82\%} (D1)
            (2, -3) coordinate (D2) circle (1pt)
            (C1) -- node[above, sloped] {18\%} (D2)
            (-3, -4) coordinate (E1) circle (1pt)
            (D1) -- node [above, sloped] {76\%} (E1)
            (-1, -4) coordinate (E2)
            (D1) -- node [above, sloped] {24\%} (E2)
            node [below, align=center] {rejected by \\ frozen bit}
            (1, -4) coordinate (E3)
            (D2) -- node [above, sloped] {72\%} (E3)
            node [below, align=center] {rejected by \\ frozen bit}
            (3, -4) coordinate (E4) circle (1pt)
            (D2) -- node [above, sloped] {28\%} (E4)
            (-3.5, -5) coordinate (F1)
            (E1) -- node [above, sloped] {36\%} (F1)
            node [below] {msg}
            (-2.5, -5) coordinate (F2)
            (E1) -- node [above, sloped] {64\%} (F2)
            node [below] {msg}
            (2.5, -5) coordinate (F3)
            (E4) -- node [above, sloped] {42\%} (F3)
            node [below] {msg}
            (3.5, -5) coordinate (F4)
            (E4) -- node [above, sloped] {58\%} (F4)
            node [below] {msg}
        ;
    \end{tikzpicture}
    \caption{
        The tree of posterior probabilities.  Msg stands for message.
        The usual SC decoder chooses the path greedily
        $51\% \to 38\% \to 82\% \to 76\% \to 64\%$.
        The SCS decoder, however, chooses this path with probability
        $51\% \cdot 38\% \cdot 82\% \cdot 76\% \cdot 64\%$.
    }                                                   \label{fig:tree}
\end{figure}
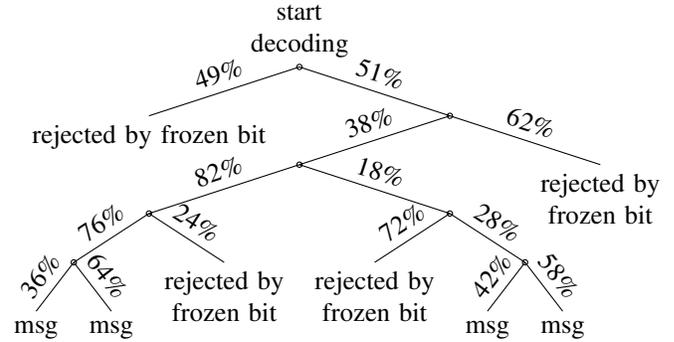

    \begin{theorem}                                  \label{thm:optimal}
        We say that a list decoder errs\footnote{ We are using the
        definition of given by Elias in \cite[abstract]{Eli91}.} if the
        correct message is not in the final list.  Let $\ell$ and $a$ be
        positive integers.  Comparing the SCS decoder with $a$ agents to
        any list decoder\footnote{ Without specifying SC, we mean
        \emph{all possible} list decoders, including SSCL \cite{HCG16},
        Fast-SSCL \cite{HCG17}, BPL \cite{EEC18}, SCAN \cite{PCB20}, and
        AE \cite{PBL21}.} with list size $\ell$, the error probability
        of the former is at most 
        \[
            \Delta \coloneqq
            \begin{cases*}
                \bigl(\frac{\ell-1}\ell\bigr)^a
                & when $a + 1 \leqslant \ell$,
                \\ \frac\ell{a+1} \bigl(\frac a{a+1}\bigr)^a
                & when $\ell \leqslant a + 1$
            \end{cases*}
        \]
        more than that of the latter.
    \end{theorem}

    \smallskip\noindent
    \textbf{Organization:}
    Section~\ref{sec:random} reviews old decoders and
    introduces new decoders.  Section~\ref{sec:optimal} proves
    Theorem~\ref{thm:optimal}.  Section~\ref{sec:natural} discusses the
    behavior of SCS for some ideal cases.  Section~\ref{sec:creative}
    discusses a generalization of SCS that is inspired by importance
    sampling.

\section{Walking on the Tree}                         \label{sec:random}

    Our idea of SCS decoder is built on the special structure of polar
    codes' SC decoder.  Let us review polar codes and SC.

\subsection{Polar codes and SC}

    A message vector $U_1^N \in \{0, 1\}^{1\times N}$ consists of
    frozen bits $(U_f)_{f\in\mathcal F}$ and information bits
    $(U_i)_{i\in\mathcal I}$.  The encoder of a polar code maps the
    message vector $U_1^N$ to a codeword
    \[
        X_1^N \coloneqq
        U_1^N \loll^{\otimes n} \in \{0, 1\}^{1\times N},
    \]
    where $n$ is some positive integer, $N \coloneqq 2^n$ is the block
    length, and $\otimes n$ means the $n$th Kronecker power.  The
    codeword $X_1^N$ is then fed into iid copies of some binary channel.
    The channel outputs $Y_1^N \in \Sigma^N$.  Here, $\Sigma$ is called
    the output alphabet.

    To decode, a circuit that is wired like a fast Fourier transformer
    will compute posterior probabilities
    \[
        \Prob( U_1{=} 0\,|\, Y_1^N )
        \text{ and }
        \Prob( U_1{=} 1\,|\, Y_1^N ).
        \label{prob:1}
    \]
    The decoder will then decide if $U_1$ is $0$ or $1$ according to the
    probabilities and whether or not $U_1$ is frozen.  In case $U_1$ is
    frozen, the correct value of $U_1$ is public information---the
    probabilities in \eqref{prob:1} will be ignored.

    Denote by $\hat U_1 \in \{0, 1\}$ the decoder's decision on $U_1$.
    The FFT-like circuit will then compute
    \[
        \Prob( U_2{=} 0\,|\, Y_1^N \hat U_1 )
        \text{ and }
        \Prob( U_2{=} 1\,|\, Y_1^N \hat U_1 ).
    \]
    The decoder will make another decision about $U_2$.
    In general, at the $(m + 1)$th step,
    \[
        \Prob( U_{m+1}{=} 0\,|\, Y_1^N \hat U_1^m )
        \text{ and }
        \Prob( U_{m+1}{=} 1\,|\, Y_1^N \hat U_1^m )
    \]
    will be computed and $\hat U_{m+1}$ will be determined.  The same
    process will be repeated until $m = N - 1$.  By that time we will
    have a full-length estimation $\hat U_1^N \in \{0, 1\}^N$.

    This is called the \emph{successive cancellation} (SC) decoder.
    See Fig.~\ref{fig:tree} for an abstract representation of SC
    as a process of choosing a root-to-leaf path.

\subsection{SCL decoder: more chances to win}

    An \emph{SC list} (SCL) decoder generalizes the SC decoder by
    maintaining two prefixes, $\dot U_1^m \in \{0, 1\}^m$ and $\ddot
    U_1^m \in \{0, 1\}^m$, in memory and choosing\footnote{ Note that
    it is possible that the first $m$ bits of $\dot U_1^{m+1}$ is $\ddot
    U_1^m$.} two from these four
    \[
        \dot U_1^m 0, \quad
        \dot U_1^m 1, \quad
        \ddot U_1^m 0, \quad
        \ddot U_1^m 1 \quad \in \{0, 1\}^{m+1}
    \]
    as $\dot U_1^{m+1}$ and $\ddot U_1^{m+1}$.  In general, the decoder
    might maintain $\ell \geqslant 1$ prefixes and choose $\ell$ out of
    $2\ell$ prefixes at every step, with $\ell$ being called the list
    size.  When $\ell = 1$, SCL falls back to SC.

    Needless to say, a larger list is Pareto-better than a smaller one
    in the sense that buying more lottery tickets increases the chance
    of matching the correct numbers.  However, it is less obvious if the
    improvement in error probability can justify the extra resources
    spent on larger lists.  In fact, it is very non-obvious due to a
    multitude of reasons that block mathematical analyses.
    \begin{itemize}
        \item Since $\dot U_1^m \neq \ddot U_1^m$, at least one of them
            is not equal to the true prefix $U_1^m$.  We do not have any
            tool that controls $\Prob( U_{m+1}{=}0 \,|\, Y_1^N \dot
            U_1^m )$ when $\dot U_1^m$ is not true.
        \item Since it is possible that $\dot U_1^m 0$ and $\dot U_1^m
            1$ will become $\dot U_1^{m+1}$ and $\ddot U_1^{m+1}$, this
            makes the prefixes in the memory highly correlated and
            worsens the tractability.
        \item Some versions of SCL even allow backtracking, i.e.,
            throwing away the trailing bits of $\dot U_1^m$ and resuming
            from a smaller $m$ \cite{YZN18, CLZ19}.
    \end{itemize}

    We try to bypass these difficulties by proposing a new list decoder
    that is more primitive than a decoder capable of sorting path
    metrics.

\subsection{SCS, a new decoder}

    The core idea of our new decoder is to replace a list by several
    agents who work independently on the prefixes.  As cooperation
    between agents is not allowed, there needs to be a way to stop them
    from all choosing the same $\hat U_1^N$: namely randomness.

    Let there be $a$ agents for some positive integer $a$.  Each agent
    will maintain a prefix $\hat U_1^m$ and choose $0$ or $1$ as $\hat
    U_{m+1}$ with probabilities
    \[
        \Prob( U_{m+1}{=}0 | Y_1^N \hat U_1^m )
        \text{ and }
        \Prob( U_{m+1}{=}1 | Y_1^N \hat U_1^m ),      \label{for:sample}
    \]
    respectively.  If at some point the agent chooses a $\hat U_{m+1}$
    that is incompatible with the frozen value, he restarts from $m =
    0$.  If the agent successfully generates a $\hat U_1^N$, he reports
    that to a manager.  We call this the \emph{SC sampling} (SCS)
    decoder with $a$ agents.  An SCS decoder succeeds if one of the
    $a$ reported messages is the correct one:  namely $U_1^N$.

    In terms of Fig.~\ref{fig:tree}, each agent will perform a random
    walk from top to bottom according to the probabilities labeled on
    the edges.  Whenever an agent hits a reject, he restarts from the
    root.  When an agent hits a leaf at the lowest level, he reports his
    path to the manager.

    Considering how this design resembles an acceptance--rejection
    sampler, we observe the following.

    \begin{lemma}                                     \label{lem:reject}
        Fix an agent.  Among all the message vectors that are compatible
        with the frozen bits, the probability that he reports a
        particular message $u_1^N$ is the posterior probability
        \[ \Prob(U_1^N{=}u_1^N \,|\, Y_1^N\text{ and frozen bits}). \]
    \end{lemma}

    \begin{IEEEproof}
        This is tautological.  To make it more obvious, consider an
        agent who insists finishing the whole vector $\hat U_1^N$ before
        checking the compatibility with the frozen bits.
    \end{IEEEproof}

\subsection{Agents restart a lot, or do they?}

    First-time readers are advised to skip this subsection.

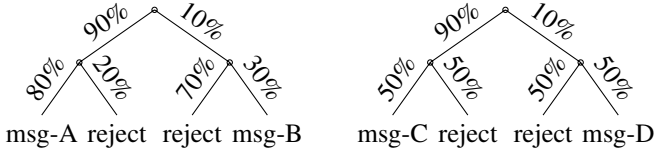
\begin{figure}
    \centering
    \begin{tikzpicture}[y=0.7cm]
        \draw 
            (0, 0) coordinate (A) circle (1pt) 
            (-1, -1) coordinate (B1) circle (1pt) 
            (A) -- node [above, sloped] {90\%} (B1)
            (1, -1) coordinate (B2) circle (1pt) 
            (A) -- node [above, sloped] {10\%} (B2)
            (-1.5, -2) coordinate (C1)
            (B1) -- node [above, sloped] {80\%} (C1)
            node [below] {msg-A}
            (-.5, -2) coordinate (C2)
            (B1) -- node [above, sloped] {20\%} (C2)
            node [below] {reject}
            (.5, -2) coordinate (C3)
            (B2) -- node [above, sloped] {70\%} (C3)
            node [below] {reject}
            (1.5, -2) coordinate (C4)
            (B2) -- node [above, sloped] {30\%} (C4)
            node [below] {msg-B}
        ;
    \end{tikzpicture}
    \hfill\hskip-1cm
    \begin{tikzpicture}[y=0.7cm]
        \draw 
            (0, 0) coordinate (A) circle (1pt) 
            (-1, -1) coordinate (B1) circle (1pt) 
            (A) -- node [above, sloped] {90\%} (B1)
            (1, -1) coordinate (B2) circle (1pt) 
            (A) -- node [above, sloped] {10\%} (B2)
            (-1.5, -2) coordinate (C1)
            (B1) -- node [above, sloped] {50\%} (C1)
            node [below] {msg-C}
            (-.5, -2) coordinate (C2)
            (B1) -- node [above, sloped] {50\%} (C2)
            node [below] {reject}
            (.5, -2) coordinate (C3)
            (B2) -- node [above, sloped] {50\%} (C3)
            node [below] {reject}
            (1.5, -2) coordinate (C4)
            (B2) -- node [above, sloped] {50\%} (C4)
            node [below] {msg-D}
        ;
    \end{tikzpicture}
    \caption{
        Left: Agent must restart.  Right: Agent might default to the
        compatible branch.
    }                                                 \label{fig:accept}
\end{figure}

    When an agent encounters a frozen bit, it is very crucial that he
    restarts from the root instead of choosing the compatible branch.
    An example is on the left-hand side of Fig.~\ref{fig:accept}: the
    desired likelihood ratio between messages A and B is $90\% \cdot
    80\% : 10\% \cdot 30\%$.  But an agent that $100\%$-chooses the
    compatible branch when he foresees a rejection will land at A and B
    with ratio $90\% : 10\%$.

    That raises a concern: Since any frozen bit can reset an agent,
    wouldn't it be very difficult to reach the bottom of the tree?  We
    argue that the concern is less severe than it seems: Most of the
    frozen bits correspond to channels where the choice between branches
    is almost 50--50, as demonstrated on the right-hand side of
    Fig.~\ref{fig:accept}.  In this case, defaulting to the compatible
    branch does not alter the likelihood ratio.

\section{How Good is SCS?}                           \label{sec:optimal}

    We now go through our proof of Theorem~\ref{thm:optimal}.  The
    theorem states that the probability that the correct message is not
    in the final list of $a$ guesses made by our SCS decoder is at most
    \[
        \Delta \coloneqq
        \begin{cases*}
            \bigl(\frac{\ell-1}\ell\bigr)^a
            & when $a + 1 \leqslant \ell$,
            \\ \frac\ell{a+1} \bigl(\frac a{a+1}\bigr)^a
            & when $\ell \leqslant a + 1$
        \end{cases*}
    \]
    plus a similarly-defined probability for any decoder with list size
    $\ell$.

\subsection{Proof of Theorem~\ref{thm:optimal}}

    To proceed, we fix the channel outputs to be some $y_1^N \in
    \Sigma^N$.  If we can prove the desired $\Delta$ when conditioning
    on an arbitrary $y_1^N$, then the same $\Delta$ holds without
    conditioning on the outputs.

    Fix the channel outputs $y_1^N \in \Sigma^N$.  Let $f\colon \{0,
    1\}^N \to [0, 1]$ be the pmf of the posterior distribution of the
    messages.  That is, for any $u_1^N \in \{0, 1\}^N$,
    \[
        f(u_1^N) \coloneqq
        \Prob( U_1^N{=}u_1^N \,|\, y_1^N \text{ and frozen bits}).
    \]
    Then, for any $u_1^N$ compatible with the frozen bits, the
    probability that none of the $a$ agents reports $u_1^N$ is $(1 -
    f(u_1^N))^a$.  The error probability of SCS is
    \[
        \sum_{u_1^N} f(u_1^N) (1 - f(u_1^N))^a.          \label{sum:all}
    \]

    For convenience, we use positive integers to relabel the messages
    sorted by the probability masses; $f(1)$ is the largest among
    $f(u_1^N)$, $f(2)$ is the second largest, and so on.
    Theorem~\ref{thm:optimal} boils down to proving that
    \[
        \sum_{k=1}^\ell f(k) f(1 - f(k))^a               \label{sum:ell}
    \]
    is $\leqslant \Delta$.  This is because the optimal decoder will get
    the $\ell$ most probable messages right, and so it suffices to focus
    on the mistakes SCS will make for $k \leqslant \ell$.

    We can now prove Theorem~\ref{thm:optimal} by running a maximization
    program on \eqref{sum:ell}.

\begin{figure}
    \centering
    \begin{tikzpicture}
        \draw [->] (0, 0) -- (8, 0) node [right] {$z$};
        \draw [->] (0, 0) -- (0, 2);
        \draw
            plot [domain=0:8, samples=100]
            (\x, {(8 - \x)^3 * \x / 256})
            (6, .5) node [rotate=-10] {$z (1 - z)^a$}
        ;
        \draw
            (2, 27/16) circle (0.5pt) circle (1pt) node [above]
            {$\bigl( \frac1{a+1},
            \frac1{a+1}\bigl(\frac a{a+1}\bigr)^a \bigr)$}
            -- node [above, pos=.7] {$g(z)$} (8, 27/16)
        ;
    \end{tikzpicture}
    \caption{
        The functions $z (1 - z)^a$ and a concave increasing upper bound
        $g(z)$.
    }                                                 \label{fig:jensen}
\end{figure}
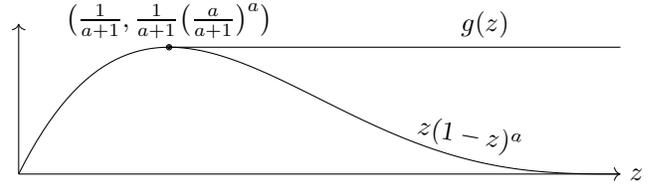

    \begin{IEEEproof}[Proof of $\eqref{sum:ell} \leq \Delta$]
        Consider the function $z (1 - z)^a$ over $z \in [0, 1]$.  This
        Bernstein polynomial takes its maximum at $z = 1/(a + 1)$.
        Therefore,
        \begin{align}
            f(k) f(1 - f(k))^a
            &\leqslant z (1 - z)^a \Bigr|_{z=1/(a+1)}      \label{ine:0}
            \\& = \frac{1}{a + 1}
                \Bigl( \frac a{a + 1} \Bigr)^a          \label{ine:stop}
            \\& = \frac1a \Bigl( \frac a{a + 1} \Bigr)^{a+1}
            < \frac1{ae}.
        \end{align}
        Now \eqref{sum:ell} sums $\ell$ terms, so \eqref{sum:ell} is
        less than $\ell/ae$, proving the claim in the abstract.  If we
        stop at \eqref{ine:stop}, we prove $\eqref{sum:ell} \leqslant
        \Delta$ for the case $\ell \leqslant a + 1$.
        
        To prove $\eqref{sum:ell} \leqslant \Delta$ for the case $a + 1
        \leqslant \ell$, we need one more thing: that $f$ is a pmf and
        hence $f(1) + f(2) + \dotsb + f(\ell) \leqslant 1$.  In order to
        apply Jensen's inequality, we define
        \[
            g(z) \coloneqq
            \cas{
                z (1 - z)^a & when $z \leqslant \frac1{a+1}$,
                \\ \frac1{a+1} \bigl(\frac a{a+1}\bigr)^a
                & when $z \geqslant \frac1{a+1}$.
            }
        \]
        With the help of Fig.~\ref{fig:jensen}, we infer
        \begin{align}
            \eqref{sum:ell}
            & \leqslant \sum_{k=1}^\ell g(f(k))            \label{ine:1}
            \\& \leqslant \ell g\Bigl(
                    \raisebox{-2pt}{$\dfrac{\sum_{k=1}^\ell f(k)}\ell$}
                \Bigr)                                     \label{ine:2}
            \\& \leqslant \ell g\Bigl( \frac1\ell \Bigr)   \label{ine:3}
            \\& = \Bigl( \frac{\ell - 1}\ell \Bigr)^a.     \label{ine:4}
        \end{align}
        \eqref{ine:1}, \eqref{ine:2}, and \eqref{ine:3} use that $g(z)$
        is an upper bound of $z (1 - z)^a$, convex, and monotonically
        increasing, respectively; and \eqref{ine:4} uses the assumption
        $a + 1 \leqslant \ell$.  This finishes the proof of
        $\eqref{sum:ell} \leqslant \Delta$ for both parameter regions,
        yielding Theorem~\ref{thm:optimal}.
    \end{IEEEproof}

\subsection{Is Theorem~\ref{thm:optimal} tight?}

    We give two concrete examples of $f$ that witness the optimality of
    the gap term $\Delta$.
    
    For $a + 1 \leqslant \ell$,
    consider the uniform distribution on $[\ell]$:
    \[
        f(k) \coloneqq \cas{
            \frac1\ell & for $k = 1, \dotsc, \ell$,
            \\ 0 & for larger $k$.
        }                                                  \label{pmf:1}
    \]
    This $f$ reflects a situation that all but $\ell$ messages are
    impossible given the channel outputs.  In this case the optimal list
    decoder will find these $\ell$ messages and enjoys error probability
    $0$.  For our SCS decoder, it can be observed that inequalities
    \eqref{ine:1}, \eqref{ine:2}, and \eqref{ine:3} are all saturated.
    So the gap term $\Delta$ cannot be improved.

    For $\ell \leqslant a + 1$, let $M$ be a very large number and
    \[
        \! f(k) \coloneqq \cas{
            \frac1{a + 1} & for $k = 1, \dotsc, \ell$,
            \\ \frac1M
            & for $k = \ell + 1, \dotsc,
            \ell + M\bigl(1 - \frac\ell{a + 1}\bigr)$.
        } \!                                               \label{pmf:2}
    \]
    This $f$ reflects a situation where some masses are on the top
    $\ell$ messages and the rest of the masses spread out.  The optimal
    list decoder can only make $\ell$ guesses so the error probability
    is $1 - \ell/(a + 1)$.  For our SCS decoder, one observes that
    \eqref{ine:0} is saturated.  That is to say, if the true message is
    among the top $\ell$ messages, SCS underperforms the optimal decoder
    by $\Delta$.  On the other hand, when the true message is not among
    the top $\ell$ messages, its contribution to the error probability
    of SCS is
    \[
        \sum_{k=\ell+1}^{\ell+M(1-\ell/(a+1))}
        \frac1M \Bigl(1 - \frac1M\Bigr)^a
        = \Bigl(1 - \frac\ell{a + 1}\Bigr) \Bigl(1 - \frac1M\Bigr)^a.
    \]
    The power term converges to $1$ as $M \to \infty$.  So eventually
    SCS errs with probability $1 - \ell/(a + 1)$, which is the same as
    the optimal list decoder.

\section{SCS and Natural PMFs}                       \label{sec:natural}

    In the previous section we argue that Theorem~\ref{thm:optimal} is
    tight by constructing adversarial pmfs \eqref{pmf:1} and
    \eqref{pmf:2}.  For applications such as wireless communication,
    this is unlikely to be the case.  In this section, we discuss the
    performance of SCS when $f$ is ``natural''.  By natural we mean
    geometric distributions and zeta distributions.  (Cf.\ (3) and (8)
    of \cite{MHU15} for supporting evidence.)

\subsection{Geometric distributions}

    We begin with
    \[ f(k) \coloneqq (1 - q) q^{k-1}, \]
    i.e., $f$ is the pmf of the geometric distribution with success rate
    $1 - q$.  For the optimal list decoder with list size $\ell$, its
    error probability is $f$'s tail mass after $k > \ell$, which is
    \[ \sum_{k=\ell+1}^\infty f(k) = q^\ell. \]

    For SCS, recall \eqref{sum:all}: the error probability is
    \[
        \sum_{k=1}^\infty
        f(k) (1 - f(k))^a
        = \sum_{k=1}^\infty
        (1 - q) q^{k-1} \bigl(1 - (1 - q) q^{k-1} \bigr)^a.
    \]
    We think it would be good to have alternative expressions.  For
    that, we need some preparation for the Taylor expansions of $z (1 -
    z)^a$.

    \begin{proposition}                               \label{pro:taylor}
        The odd-order Taylor expansions of $z (1 - z)^a$ at $z = 0$
        are upper bounds on the interval $z \in [0, 1]$.
        The even-order Taylor expansions of $z (1 - z)^a$ at $z = 0$
        are lower bounds on the interval $z \in [0, 1]$.
    \end{proposition}

    \begin{IEEEproof}
        The sequence of successive derivatives of $(1 - z)^a$ reads $(1
        - z)^a$, $-a (1 - z)^{a-1}$, $a (a - 1) (1 - z)^{a-2}$, $-a (a -
        1) \* (a - 2) (1 - z)^{a-3}$, and so on.  Their signs alternate.
        Because the remainder term of the $t$th-order Taylor expansion
        is its $(t + 1)$th order derivative evaluated at somewhere
        within $[0, 1]$, the $t$th-order Taylor expansion of $(1 - z)^a$
        is a lower bound if $t$ is odd, upper bound if $t$ is even.
        Multiplied by $z$, the Taylor expansions of $z (1 - z)^a$ also
        alternate between upper bounds and lower bounds.
    \end{IEEEproof}

    Proposition~\ref{pro:taylor} provides an elastic framework to upper
    and lower bound the error probability.  For instance, the
    first-order Taylor expansion of $z (1 - z)^a$ at $z = 0$ is just
    $z$.  So the error probability is
    \[ \leqslant \sum_{k=1}^\infty f(k) = 1, \]
    which is vacuous.  The second-order Taylor expansion of $z (1 -
    z)^a$ at $z = 0$ is $z - az^2$, implying that the error probability
    is
    \begin{align}
        \geqslant \sum_{k=1}^\infty f(k) - af(k)^2
        & = 1 - a \sum_{k=1}^\infty ((1 - q) q^{k-1})^2
        \\& = 1 - a \frac{(1 - q)^2}{1 - q^2}.
    \end{align}
    Now apply the third-order bound $z (1 - z)^a \leqslant z - az^2 +
    \binom a2 z^3$, we obtain that the error probability of SCS is
    \begin{align}
        & \leqslant \sum_{k=1}^\infty f(k) - af(k)^2 + \binom a2 f(k)^3
        \\& = 1 - a \frac{(1 - q)^2}{1 - q^2}
        + \binom a2 \frac{(1 - q)^3}{1 - q^3}.
    \end{align}
    It is not hard to see that the $t$th term of the expansion of the
    error probability is $\binom at (1 - q)^t / (1 - q^t)$; it vanishes
    after $t > a$.

\subsection{Zeta distributions}

    Using an almost exact argument, we can express the error probability
    corresponding to a zeta distribution
    \[ f(k) \coloneqq \frac1{\zeta(s) k^s} \]
   as (note that this is a finite sum)
    \[
        \binom a0 \frac{\zeta(s)}{\zeta(s)}
        - \binom a1 \frac{\zeta(2s)}{\zeta(s)^2}
        + \binom a2 \frac{\zeta(3s)}{\zeta(s)^3}
        - \binom a3 \frac{\zeta(4s)}{\zeta(s)^4}
        + \dotsb
    \]
    and assert that its truncations alternate between upper bounds and
    lower bounds.

\subsection{Limiting behavior when there are many agents}

    Earlier this section, we see alternative expressions of the error
    probabilities when $f$ is natural.  In this subsection, we study the
    asymptotic behavior when $a$ goes to infinity.

    We demonstrate our point using a geometric distribution
    \[ f(k) \coloneqq (1 - q) q^{k-1}. \]
    We observe that the behavior of $f(k) (1 - f(k))^a$ can be divided
    into three cases.
    \begin{itemize}
        \item Large $k$: for a $k$ such that $f(k) < 1/a$,
            the first-order upper bound $f(k) (1 - f(k))^a < f(k)$
            forms a geometric sequence as $k$ increases.
            These $k$'s contribute $O(1/a(1 - q))$ to the error.
        \item Mediocre $k$: for $f(k)$ between $1/a$ and $2/a$,
            we upper bound $f(k) (1 - f(k))^a$ by the maximum value
            of $z (1 - z)^a$, which is $O(1/a)$.
            The number of such $k$'s is $-\log_q(2) < O(1/(1 - q))$.
            They contribute $O(1/a(1 - q))$ to the error.
        \item Small $k$: for a $k$ such that $f(k) > 2/a$,
            upper bound the power term as $f(k) (1 - f(k))^a
            < f(k) \exp(-a f(k)) < f(k) / a^2 f(k)^2 = 1 / a^2 f(k)$.
            This forms a geometric sequence as $k$ decreases.
            These $k$'s contribute $O(1/a(1 - q))$ to the error.
    \end{itemize}

    \begin{proposition}                                  \label{pro:geo}
        If, regardless of the channel outputs, the posterior
        probabilities of the messages follow a geometric distribution
        with success rate $1 - q$, then the error probability of the SCS
        decoder with $a$ agents is $O(1/a(1 - q))$.
    \end{proposition}

    Compare Proposition~\ref{pro:geo} with Theorem~\ref{thm:optimal}.
    The upper bound on the error probability given by the theorem is
    $\Delta = O(\ell/a)$ plus the error of the optimal decoder $q^\ell$.
    We can freely pick $\ell$ and the nearly-optimal choice $\ell
    \coloneqq -\log_q(a)$ results in an inferior bound $O(\log(a) / a(1
    - q))$.

\section{Customize the Temperature}                 \label{sec:creative}

    In this section, we propose an avenue to improve the behavior of
    SCS.  We call it an \emph{SC importance sampling} (SCIS) decoder for
    that it samples the messages using a twisted distribution.  The
    twisting is controlled by a free parameter $\beta > 0$ that is
    sometimes called the \emph{inverse temperature}.

\subsection{Twist the distribution}

    In \eqref{for:sample}, an SCS agent chooses the next bit $\hat
    U_{m+1}$ based on the posterior probabilities
    \[
        \Prob( U_{m+1}{=} 0\,|\, Y_1^N \hat U_1^m )
        \text{ and }
        \Prob( U_{m+1}{=} 1\,|\, Y_1^N \hat U_1^m ),
    \]
    An SCIS agent, however, chooses the next bit using the ratio
    \[
        \Prob( U_{m+1}{=} 0\,|\, Y_1^N \hat U_1^m )^\beta
        :
        \Prob( U_{m+1}{=} 1\,|\, Y_1^N \hat U_1^m )^\beta.
    \]
    When $\beta = 1$, SCIS falls back to SCS.  When $\beta < 1$, an SCIS
    agent is ``warmer'' and willing to invest in less probable prefixes
    just to avoid getting the same answer as other agents.

    What is beautiful about SCIS is that raising to the power of $\beta$
    commutes with chaining of the conditional probabilities, hence we
    have the following generalization of Lemma~\ref{lem:reject}.

    \begin{lemma}
        Fix an agent.  Among all the message vectors that are compatible
        with the frozen bits, the probability that he reports a
        particular message $u_1^N$ is
        \[
            \frac1{Z(\beta)} \Prob(
                U_1^N{=}u_1^N \,|\, Y_1^N\text{ and frozen bits}
            )^\beta,
        \]
        for some normalizing constant $Z(\beta) > 0$ that makes the
        probabilities sum to $1$.
    \end{lemma}

\subsection{Does twisting help?}

    We continue using $f$ to represent the pmf.  We use $f_\beta$ to
    represent the pmf that is proportional to $f^\beta$.  The error
    probability of the SCIS decoder with $a$ agents is then
    \[ \sum_{k=1}^\infty f(k) (1 - f_\beta(k))^a. \]
    When $f(k)$ is about $\Theta(1/a)$, $(1 - f(k))^a$ is roughly a
    constant, meaning that the error probability of SCS is at least
    $\Omega(1/a)$.  Now that this term becomes $(1 - f_\beta(k))^a$,
    chances are that $f_\beta(k)$ is significantly greater than $f(k)$
    and that $(1 - f_\beta(k))^a$ is significantly smaller.  Should that
    happen, this $k$ contributes a very tiny $f(k) (1 - f_\beta(k))^a$
    to the error probability.

\begin{figure}
    \centering
    \begin{tikzpicture}
        \begin{axis}[xlabel=$\beta$, ylabel=error probability]
            \addplot [red ] table [x=x, y=y1  , mark=none] {\tablebeta};
            \addplot [teal] table [x=x, y=y2  , mark=none] {\tablebeta};
            \addplot [blue] table [x=x, y=y4  , mark=none] {\tablebeta};
            \addplot [red ] table [x=x, y=y8  , mark=none] {\tablebeta};
            \addplot [teal] table [x=x, y=y16 , mark=none] {\tablebeta};
            \addplot [blue] table [x=x, y=y32 , mark=none] {\tablebeta};
            \addplot [red ] table [x=x, y=y64 , mark=none] {\tablebeta};
            \addplot [teal] table [x=x, y=y128, mark=none] {\tablebeta};
            \addplot [blue] table [x=x, y=y256, mark=none] {\tablebeta};
        \end{axis}
    \end{tikzpicture}
    \caption{
        Error probability as a function of $\beta$.  We use geometric
        distribution with success rate $1 - q = 0.1$.  One curve for a
        fixed $a$.  From top to bottom: $a = 1, 2, 4, \dotsc, 256$.
    }                                                   \label{fig:beta}
\end{figure}
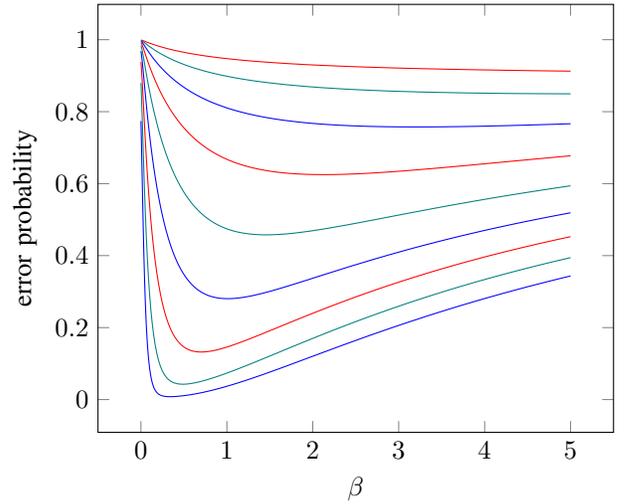

    Fig.~\ref{fig:beta} presents simulations that support the heuristic
    above.

    We conclude this paper with a remark that not all agents need to use
    the same $\beta$.  For instance, if the first agent uses $\beta =
    \infty$ and the others use finite $\beta$, then first agent will
    recover the behavior of the SC decoder while the others remain SCIS.
    This way, SCIS with customized assignments of $\beta$ will strictly
    outperform the SC decoder.  We, however, cannot determine if there
    is any real benefit in it.

\pgfplotstableread{
x      y1     y2     y4     y8     y16    y32    y64    y128   y256
0.0    0.999  0.998  0.996  0.992  0.9841 0.9685 0.938  0.8798 0.774  
0.0005 0.999  0.9979 0.9959 0.9918 0.9837 0.9677 0.9364 0.8768 0.7688 
0.002  0.9989 0.9978 0.9956 0.9912 0.9824 0.9651 0.9314 0.8676 0.7527 
0.0046 0.9987 0.9975 0.995  0.99   0.9801 0.9606 0.9227 0.8515 0.725  
0.0082 0.9985 0.997  0.9941 0.9882 0.9766 0.9537 0.9096 0.8274 0.6846 
0.0128 0.9982 0.9964 0.9928 0.9857 0.9717 0.9441 0.8914 0.7946 0.6313 
0.0184 0.9978 0.9956 0.9911 0.9824 0.9651 0.9313 0.8674 0.7523 0.566  
0.025  0.9972 0.9945 0.989  0.9781 0.9566 0.9151 0.8374 0.7013 0.4919 
0.0326 0.9966 0.9931 0.9863 0.9728 0.9463 0.8956 0.802  0.6433 0.4139 
0.0413 0.9958 0.9915 0.9832 0.9666 0.9343 0.873  0.7622 0.581  0.3377 
0.051  0.9949 0.9898 0.9796 0.9596 0.9209 0.8481 0.7193 0.5175 0.2681 
0.0617 0.9939 0.9878 0.9757 0.952  0.9063 0.8213 0.6747 0.4554 0.2079 
0.0735 0.9928 0.9856 0.9715 0.9437 0.8907 0.7933 0.6295 0.3967 0.1581 
0.0862 0.9916 0.9833 0.967  0.935  0.8743 0.7645 0.5848 0.3426 0.1184 
0.1    0.9904 0.9809 0.9622 0.9259 0.8574 0.7353 0.541  0.2937 0.0876 
0.1148 0.9892 0.9784 0.9573 0.9165 0.84   0.7058 0.4989 0.2504 0.0643 
0.1306 0.9878 0.9758 0.9522 0.9067 0.8223 0.6765 0.4587 0.2125 0.0472 
0.1474 0.9865 0.9731 0.947  0.8968 0.8044 0.6476 0.4208 0.1798 0.0347 
0.1653 0.9851 0.9703 0.9416 0.8866 0.7864 0.6192 0.3854 0.152  0.0258 
0.1842 0.9836 0.9675 0.9361 0.8764 0.7684 0.5915 0.3526 0.1287 0.0196 
0.2041 0.9821 0.9646 0.9305 0.866  0.7505 0.5647 0.3225 0.1093 0.0153 
0.225  0.9807 0.9617 0.9249 0.8556 0.7328 0.5389 0.2951 0.0935 0.0123 
0.2469 0.9791 0.9587 0.9192 0.8453 0.7153 0.5143 0.2703 0.0806 0.0104 
0.2699 0.9776 0.9558 0.9136 0.8349 0.6982 0.4909 0.2481 0.0704 0.0092 
0.2938 0.9761 0.9528 0.9079 0.8247 0.6815 0.4687 0.2284 0.0623 0.0084 
0.3188 0.9745 0.9498 0.9022 0.8145 0.6653 0.4479 0.2111 0.056  0.0081 
0.3449 0.973  0.9468 0.8966 0.8045 0.6496 0.4284 0.1959 0.0512 0.008  
0.3719 0.9714 0.9438 0.891  0.7947 0.6344 0.4103 0.1828 0.0477 0.0082 
0.4    0.9699 0.9408 0.8854 0.7851 0.6199 0.3935 0.1716 0.0453 0.0085 
0.429  0.9684 0.9379 0.88   0.7757 0.6059 0.3781 0.1621 0.0437 0.0091 
0.4591 0.9669 0.9349 0.8746 0.7665 0.5926 0.364  0.1542 0.0429 0.0098 
0.4903 0.9654 0.9321 0.8693 0.7576 0.5799 0.3512 0.1477 0.0426 0.0106 
0.5224 0.9639 0.9292 0.8641 0.7489 0.5679 0.3397 0.1425 0.0429 0.0116 
0.5556 0.9624 0.9264 0.859  0.7405 0.5566 0.3293 0.1385 0.0437 0.0127 
0.5897 0.9609 0.9236 0.854  0.7325 0.546  0.3202 0.1356 0.0449 0.014  
0.6249 0.9595 0.9209 0.8492 0.7247 0.536  0.3121 0.1336 0.0464 0.0154 
0.6612 0.9581 0.9183 0.8444 0.7172 0.5268 0.3051 0.1325 0.0483 0.0171 
0.6984 0.9567 0.9157 0.8398 0.7101 0.5181 0.2991 0.1321 0.0504 0.0188 
0.7367 0.9553 0.9131 0.8354 0.7033 0.5102 0.2941 0.1324 0.0529 0.0208 
0.7759 0.954  0.9106 0.831  0.6967 0.5029 0.2899 0.1333 0.0556 0.0229 
0.8162 0.9527 0.9082 0.8268 0.6905 0.4962 0.2865 0.1347 0.0585 0.0252 
0.8576 0.9514 0.9058 0.8228 0.6847 0.4901 0.284  0.1367 0.0617 0.0276 
0.8999 0.9501 0.9035 0.8189 0.6791 0.4847 0.2821 0.1391 0.0651 0.0303 
0.9433 0.9489 0.9012 0.8151 0.6738 0.4798 0.281  0.1419 0.0687 0.0331 
0.9877 0.9477 0.899  0.8114 0.6689 0.4754 0.2804 0.145  0.0725 0.0361 
1.0331 0.9465 0.8969 0.8079 0.6642 0.4716 0.2804 0.1485 0.0765 0.0393 
1.0795 0.9454 0.8948 0.8046 0.6599 0.4683 0.281  0.1522 0.0807 0.0426 
1.1269 0.9442 0.8928 0.8014 0.6558 0.4655 0.282  0.1563 0.085  0.0461 
1.1754 0.9431 0.8908 0.7983 0.6521 0.4632 0.2835 0.1605 0.0896 0.0498 
1.2249 0.9421 0.8889 0.7953 0.6486 0.4613 0.2854 0.165  0.0942 0.0537 
1.2754 0.941  0.8871 0.7925 0.6454 0.4599 0.2877 0.1697 0.0991 0.0577 
1.3269 0.94   0.8853 0.7898 0.6425 0.4588 0.2903 0.1746 0.1041 0.0619 
1.3795 0.939  0.8836 0.7873 0.6398 0.4581 0.2932 0.1796 0.1092 0.0662 
1.433  0.938  0.8819 0.7848 0.6374 0.4578 0.2964 0.1848 0.1144 0.0707 
1.4876 0.9371 0.8803 0.7825 0.6352 0.4578 0.2999 0.1901 0.1198 0.0754 
1.5432 0.9362 0.8788 0.7803 0.6332 0.4582 0.3036 0.1955 0.1253 0.0801 
1.5998 0.9353 0.8773 0.7783 0.6315 0.4588 0.3075 0.2011 0.1309 0.085  
1.6575 0.9344 0.8758 0.7763 0.63   0.4597 0.3116 0.2067 0.1366 0.0901 
1.7162 0.9335 0.8744 0.7745 0.6287 0.4609 0.3158 0.2125 0.1424 0.0952 
1.7758 0.9327 0.8731 0.7728 0.6277 0.4623 0.3203 0.2183 0.1483 0.1005 
1.8365 0.9319 0.8718 0.7711 0.6268 0.464  0.3248 0.2242 0.1542 0.1059 
1.8983 0.9311 0.8705 0.7696 0.6261 0.4659 0.3294 0.2301 0.1602 0.1114 
1.961  0.9303 0.8693 0.7682 0.6256 0.4679 0.3342 0.2361 0.1663 0.117  
2.0248 0.9296 0.8682 0.7669 0.6253 0.4702 0.3391 0.2422 0.1724 0.1226 
2.0896 0.9289 0.867  0.7657 0.6251 0.4726 0.344  0.2482 0.1786 0.1284 
2.1554 0.9282 0.866  0.7646 0.6251 0.4751 0.349  0.2543 0.1849 0.1342 
2.2222 0.9275 0.865  0.7636 0.6252 0.4778 0.354  0.2605 0.1912 0.1401 
2.2901 0.9268 0.864  0.7626 0.6255 0.4807 0.3591 0.2666 0.1975 0.1461 
2.3589 0.9262 0.863  0.7618 0.6259 0.4836 0.3643 0.2728 0.2038 0.1521 
2.4288 0.9255 0.8621 0.761  0.6265 0.4867 0.3695 0.279  0.2102 0.1582 
2.4997 0.9249 0.8613 0.7604 0.6271 0.4898 0.3747 0.2852 0.2166 0.1643 
2.5717 0.9243 0.8605 0.7598 0.6279 0.493  0.3799 0.2914 0.223  0.1705 
2.6446 0.9237 0.8597 0.7592 0.6288 0.4964 0.3851 0.2975 0.2294 0.1767 
2.7186 0.9232 0.8589 0.7588 0.6299 0.4997 0.3904 0.3037 0.2358 0.1829 
2.7936 0.9226 0.8582 0.7584 0.631  0.5032 0.3956 0.3099 0.2422 0.1892 
2.8696 0.9221 0.8575 0.7581 0.6322 0.5067 0.4009 0.316  0.2486 0.1955 
2.9466 0.9215 0.8569 0.7579 0.6335 0.5102 0.4061 0.3221 0.255  0.2018 
3.0247 0.921  0.8563 0.7577 0.6348 0.5138 0.4113 0.3282 0.2614 0.2081 
3.1038 0.9205 0.8557 0.7576 0.6363 0.5174 0.4166 0.3343 0.2678 0.2144 
3.1839 0.9201 0.8552 0.7576 0.6378 0.521  0.4218 0.3403 0.2742 0.2207 
3.265  0.9196 0.8546 0.7576 0.6394 0.5247 0.4269 0.3463 0.2805 0.227  
3.3471 0.9191 0.8542 0.7576 0.6411 0.5284 0.4321 0.3523 0.2869 0.2334 
3.4303 0.9187 0.8537 0.7578 0.6428 0.5321 0.4372 0.3583 0.2931 0.2397 
3.5144 0.9182 0.8533 0.7579 0.6445 0.5358 0.4423 0.3642 0.2994 0.246  
3.5996 0.9178 0.8529 0.7582 0.6464 0.5395 0.4474 0.37   0.3057 0.2523 
3.6858 0.9174 0.8525 0.7584 0.6482 0.5432 0.4524 0.3759 0.3119 0.2586 
3.7731 0.917  0.8521 0.7588 0.6501 0.5469 0.4574 0.3817 0.318  0.2648 
3.8613 0.9166 0.8518 0.7591 0.6521 0.5506 0.4624 0.3874 0.3242 0.2711 
3.9506 0.9162 0.8515 0.7595 0.6541 0.5543 0.4674 0.3931 0.3303 0.2773 
4.0409 0.9159 0.8512 0.76   0.6561 0.558  0.4723 0.3988 0.3363 0.2835 
4.1322 0.9155 0.851  0.7605 0.6581 0.5617 0.4771 0.4044 0.3423 0.2896 
4.2246 0.9151 0.8507 0.761  0.6602 0.5653 0.4819 0.4099 0.3483 0.2958 
4.3179 0.9148 0.8505 0.7616 0.6623 0.569  0.4867 0.4154 0.3542 0.3019 
4.4123 0.9145 0.8503 0.7622 0.6644 0.5726 0.4915 0.4209 0.3601 0.3079 
4.5077 0.9141 0.8502 0.7628 0.6665 0.5762 0.4961 0.4263 0.3659 0.314  
4.6041 0.9138 0.85   0.7635 0.6687 0.5798 0.5008 0.4317 0.3717 0.3199 
4.7016 0.9135 0.8499 0.7642 0.6708 0.5834 0.5054 0.437  0.3775 0.3259 
4.8    0.9132 0.8498 0.7649 0.673  0.5869 0.51   0.4423 0.3832 0.3318 
4.8995 0.9129 0.8497 0.7657 0.6752 0.5905 0.5145 0.4475 0.3888 0.3377 
5.0    0.9126 0.8496 0.7665 0.6774 0.594  0.519  0.4526 0.3944 0.3435 
}\tablebeta

\section{Conclusion}

    We propose SCS decoder to help understand list decoding.  Our
    decoder samples the messages using the posterior distribution,
    making it easy to analyze.  We also propose SCIS decoder that
    generalizes SCS with a free parameter $\beta > 0$.  Adjusting
    $\beta$ sometimes improves SCS.  The simple structures of SCS and
    SCIS allow immediate generalization to list-decoding
    polarization-adjusted convolutional (PAC) codes \cite{RBV21, RoV21,
    YFV21} and, in general, any successively decoded codes.

\section{Acknowledgment}

    Research supported in part by
    NSF grant CCF-2210823 and a Simons Investigator Award.

\IEEEtriggeratref{14}
\bibliographystyle{IEEEtran}
\bibliography{SCPDecoder-29}

\begin{thebibliography}{10}
\providecommand{\url}[1]{#1}
\csname url@samestyle\endcsname
\providecommand{\newblock}{\relax}
\providecommand{\bibinfo}[2]{#2}
\providecommand{\BIBentrySTDinterwordspacing}{\spaceskip=0pt\relax}
\providecommand{\BIBentryALTinterwordstretchfactor}{4}
\providecommand{\BIBentryALTinterwordspacing}{\spaceskip=\fontdimen2\font plus
\BIBentryALTinterwordstretchfactor\fontdimen3\font minus \fontdimen4\font\relax}
\providecommand{\BIBforeignlanguage}[2]{{%
\expandafter\ifx\csname l@#1\endcsname\relax
\typeout{** WARNING: IEEEtran.bst: No hyphenation pattern has been}%
\typeout{** loaded for the language `#1'. Using the pattern for}%
\typeout{** the default language instead.}%
\else
\language=\csname l@#1\endcsname
\fi
#2}}
\providecommand{\BIBdecl}{\relax}
\BIBdecl

\bibitem{Ari09}
E.~Arikan, ``Channel {{Polarization}}: {{A Method}} for {{Constructing Capacity-Achieving Codes}} for {{Symmetric Binary-Input Memoryless Channels}},'' \emph{IEEE Transactions on Information Theory}, vol.~55, no.~7, pp. 3051--3073, Jul. 2009.

\bibitem{STA09}
\BIBentryALTinterwordspacing
E.~Sasoglu, E.~Telatar, and E.~Arikan, ``Polarization for arbitrary discrete memoryless channels,'' in \emph{2009 {{IEEE Information Theory Workshop}}}.\hskip 1em plus 0.5em minus 0.4em\relax Taormina: IEEE, Oct. 2009, pp. 144--148. [Online]. Available: \url{http://ieeexplore.ieee.org/document/5351487/}
\BIBentrySTDinterwordspacing

\bibitem{MEL16}
\BIBentryALTinterwordspacing
H.~Mahdavifar, M.~{El-Khamy}, J.~Lee, and I.~Kang, ``Achieving the {{Uniform Rate Region}} of {{General Multiple Access Channels}} by {{Polar Coding}},'' \emph{IEEE Transactions on Communications}, vol.~64, no.~2, pp. 467--478, Feb. 2016. [Online]. Available: \url{http://ieeexplore.ieee.org/document/7352336/}
\BIBentrySTDinterwordspacing

\bibitem{GAG15}
\BIBentryALTinterwordspacing
N.~Goela, E.~Abbe, and M.~Gastpar, ``Polar {{Codes}} for {{Broadcast Channels}},'' \emph{IEEE Transactions on Information Theory}, vol.~61, no.~2, pp. 758--782, Feb. 2015. [Online]. Available: \url{https://ieeexplore.ieee.org/document/6975233}
\BIBentrySTDinterwordspacing

\bibitem{MaV11}
\BIBentryALTinterwordspacing
H.~Mahdavifar and A.~Vardy, ``Achieving the {{Secrecy Capacity}} of {{Wiretap Channels Using Polar Codes}},'' \emph{IEEE Transactions on Information Theory}, vol.~57, no.~10, pp. 6428--6443, Oct. 2011. [Online]. Available: \url{http://ieeexplore.ieee.org/document/6034749/}
\BIBentrySTDinterwordspacing

\bibitem{HoY13}
\BIBentryALTinterwordspacing
J.~Honda and H.~Yamamoto, ``Polar {{Coding Without Alphabet Extension}} for {{Asymmetric Models}},'' \emph{IEEE Transactions on Information Theory}, vol.~59, no.~12, pp. 7829--7838, Dec. 2013. [Online]. Available: \url{http://ieeexplore.ieee.org/document/6601656/}
\BIBentrySTDinterwordspacing

\bibitem{GNS19}
\BIBentryALTinterwordspacing
V.~Guruswami, P.~Nakkiran, and M.~Sudan, ``Algorithmic {{Polarization}} for {{Hidden Markov Models}},'' in \emph{{{DROPS-IDN}}/v2/Document/10.4230/{{LIPIcs}}.{{ITCS}}.2019.39}.\hskip 1em plus 0.5em minus 0.4em\relax Schloss Dagstuhl -- Leibniz-Zentrum f{\"u}r Informatik, 2019. [Online]. Available: \url{https://drops.dagstuhl.de/entities/document/10.4230/LIPIcs.ITCS.2019.39}
\BIBentrySTDinterwordspacing

\bibitem{Mah20}
\BIBentryALTinterwordspacing
H.~Mahdavifar, ``Polar {{Coding}} for {{Non-Stationary Channels}},'' \emph{IEEE Transactions on Information Theory}, vol.~66, no.~11, pp. 6920--6938, Nov. 2020. [Online]. Available: \url{https://ieeexplore.ieee.org/document/9184107/}
\BIBentrySTDinterwordspacing

\bibitem{TPF22}
\BIBentryALTinterwordspacing
I.~Tal, H.~D. Pfister, A.~Fazeli, and A.~Vardy, ``Polar {{Codes}} for the {{Deletion Channel}}: {{Weak}} and {{Strong Polarization}},'' \emph{IEEE Transactions on Information Theory}, vol.~68, no.~4, pp. 2239--2265, Apr. 2022. [Online]. Available: \url{https://ieeexplore.ieee.org/document/9654208/}
\BIBentrySTDinterwordspacing

\bibitem{TaV15}
I.~Tal and A.~Vardy, ``List {{Decoding}} of {{Polar Codes}},'' \emph{IEEE Transactions on Information Theory}, vol.~61, no.~5, pp. 2213--2226, May 2015.

\bibitem{BCL21}
V.~Bioglio, C.~Condo, and I.~Land, ``Design of {{Polar Codes}} in {{5G New Radio}},'' \emph{IEEE Communications Surveys \& Tutorials}, vol.~23, no.~1, pp. 29--40, 2021.

\bibitem{MHU15}
M.~Mondelli, S.~H. Hassani, and R.~L. Urbanke, ``Scaling {{Exponent}} of {{List Decoders With Applications}} to {{Polar Codes}},'' \emph{IEEE Transactions on Information Theory}, vol.~61, no.~9, pp. 4838--4851, Sep. 2015.

\bibitem{CoP22}
M.~C. Coskun and H.~D. Pfister, ``An {{Information-Theoretic Perspective}} on {{Successive Cancellation List Decoding}} and {{Polar Code Design}},'' \emph{IEEE Transactions on Information Theory}, vol.~68, no.~9, pp. 5779--5791, Sep. 2022.

\bibitem{HMH18}
\BIBentryALTinterwordspacing
S.~A. Hashemi, M.~Mondelli, S.~H. Hassani, C.~Condo, R.~L. Urbanke, and W.~J. Gross, ``Decoder {{Partitioning}}: {{Towards Practical List Decoding}} of {{Polar Codes}},'' \emph{IEEE Transactions on Communications}, vol.~66, no.~9, pp. 3749--3759, Sep. 2018. [Online]. Available: \url{https://ieeexplore.ieee.org/document/8353453/}
\BIBentrySTDinterwordspacing

\bibitem{FVY21}
A.~Fazeli, A.~Vardy, and H.~Yao, ``List {{Decoding}} of {{Polar Codes}}: {{How Large Should}} the {{List Be}} to {{Achieve ML Decoding}}?'' in \emph{2021 {{IEEE International Symposium}} on {{Information Theory}} ({{ISIT}})}.\hskip 1em plus 0.5em minus 0.4em\relax {Melbourne, Australia}: {IEEE}, Jul. 2021, pp. 1594--1599.

\bibitem{Eli91}
\BIBentryALTinterwordspacing
P.~Elias, ``Error-correcting codes for list decoding,'' \emph{IEEE Transactions on Information Theory}, vol.~37, no.~1, pp. 5--12, Jan./1991. [Online]. Available: \url{http://ieeexplore.ieee.org/document/61123/}
\BIBentrySTDinterwordspacing

\bibitem{HCG16}
\BIBentryALTinterwordspacing
S.~A. Hashemi, C.~Condo, and W.~J. Gross, ``Simplified {{Successive-Cancellation List}} decoding of polar codes,'' in \emph{2016 {{IEEE International Symposium}} on {{Information Theory}} ({{ISIT}})}.\hskip 1em plus 0.5em minus 0.4em\relax Barcelona, Spain: IEEE, Jul. 2016, pp. 815--819. [Online]. Available: \url{http://ieeexplore.ieee.org/document/7541412/}
\BIBentrySTDinterwordspacing

\bibitem{HCG17}
\BIBentryALTinterwordspacing
------, ``Fast {{Simplified Successive-Cancellation List Decoding}} of {{Polar Codes}},'' in \emph{2017 {{IEEE Wireless Communications}} and {{Networking Conference Workshops}} ({{WCNCW}})}.\hskip 1em plus 0.5em minus 0.4em\relax San Francisco, CA, USA: IEEE, Mar. 2017, pp. 1--6. [Online]. Available: \url{http://ieeexplore.ieee.org/document/7919044/}
\BIBentrySTDinterwordspacing

\bibitem{EEC18}
\BIBentryALTinterwordspacing
A.~Elkelesh, M.~Ebada, S.~Cammerer, and S.~Ten~Brink, ``Belief {{Propagation List Decoding}} of {{Polar Codes}},'' \emph{IEEE Communications Letters}, vol.~22, no.~8, pp. 1536--1539, Aug. 2018. [Online]. Available: \url{https://ieeexplore.ieee.org/document/8396299/}
\BIBentrySTDinterwordspacing

\bibitem{PCB20}
\BIBentryALTinterwordspacing
C.~Pillet, C.~Condo, and V.~Bioglio, ``{{SCAN List Decoding}} of {{Polar Codes}},'' in \emph{{{ICC}} 2020 - 2020 {{IEEE International Conference}} on {{Communications}} ({{ICC}})}.\hskip 1em plus 0.5em minus 0.4em\relax Dublin, Ireland: IEEE, Jun. 2020, pp. 1--6. [Online]. Available: \url{https://ieeexplore.ieee.org/document/9149202/}
\BIBentrySTDinterwordspacing

\bibitem{PBL21}
\BIBentryALTinterwordspacing
C.~Pillet, V.~Bioglio, and I.~Land, ``Polar {{Codes}} for {{Automorphism Ensemble Decoding}},'' in \emph{2021 {{IEEE Information Theory Workshop}} ({{ITW}})}.\hskip 1em plus 0.5em minus 0.4em\relax Kanazawa, Japan: IEEE, Oct. 2021, pp. 1--6. [Online]. Available: \url{https://ieeexplore.ieee.org/document/9611504/}
\BIBentrySTDinterwordspacing

\bibitem{YZN18}
\BIBentryALTinterwordspacing
Y.~Yongrun, P.~Zhiwen, L.~Nan, and Y.~Xiaohu, ``Successive {{Cancellation List Bit-flip Decoder}} for {{Polar Codes}},'' in \emph{2018 10th {{International Conference}} on {{Wireless Communications}} and {{Signal Processing}} ({{WCSP}})}.\hskip 1em plus 0.5em minus 0.4em\relax Hangzhou: IEEE, Oct. 2018, pp. 1--6. [Online]. Available: \url{https://ieeexplore.ieee.org/document/8555688/}
\BIBentrySTDinterwordspacing

\bibitem{CLZ19}
\BIBentryALTinterwordspacing
F.~Cheng, A.~Liu, Y.~Zhang, and J.~Ren, ``Bit-{{Flip Algorithm}} for {{Successive Cancellation List Decoder}} of {{Polar Codes}},'' \emph{IEEE Access}, vol.~7, pp. 58\,346--58\,352, 2019. [Online]. Available: \url{https://ieeexplore.ieee.org/document/8705213/}
\BIBentrySTDinterwordspacing

\bibitem{RBV21}
\BIBentryALTinterwordspacing
M.~Rowshan, A.~Burg, and E.~Viterbo, ``Polarization-{{Adjusted Convolutional}} ({{PAC}}) {{Codes}}: {{Sequential Decoding}} vs {{List Decoding}},'' \emph{IEEE Transactions on Vehicular Technology}, vol.~70, no.~2, pp. 1434--1447, Feb. 2021. [Online]. Available: \url{https://ieeexplore.ieee.org/document/9328621/}
\BIBentrySTDinterwordspacing

\bibitem{RoV21}
\BIBentryALTinterwordspacing
M.~Rowshan and E.~Viterbo, ``List {{Viterbi Decoding}} of {{PAC Codes}},'' \emph{IEEE Transactions on Vehicular Technology}, vol.~70, no.~3, pp. 2428--2435, Mar. 2021. [Online]. Available: \url{https://ieeexplore.ieee.org/document/9354542/}
\BIBentrySTDinterwordspacing

\bibitem{YFV21}
\BIBentryALTinterwordspacing
H.~Yao, A.~Fazeli, and A.~Vardy, ``List {{Decoding}} of {{Ar{\i}kan}}'s {{PAC Codes}},'' \emph{Entropy}, vol.~23, no.~7, p. 841, Jun. 2021. [Online]. Available: \url{https://www.mdpi.com/1099-4300/23/7/841}
\BIBentrySTDinterwordspacing

\end{thebibliography}

\end{document}